\documentclass[12pt]{article}

\usepackage{multirow}
\usepackage{verbatim}
\usepackage{apacite}

\usepackage{authblk} 
\usepackage{float} 
\usepackage{amsmath,amsthm,amssymb,bbm,mathrsfs,mathtools,xfrac} 

\usepackage{adjustbox}
\usepackage[round,authoryear]{natbib}   
\usepackage{placeins} 
\usepackage[pagebackref=true,bookmarks]{hyperref}
\hypersetup{
	unicode=false,
	pdftoolbar=true,
	pdfmenubar=true,
	pdffitwindow=false,     
	pdfstartview={FitH},    
	pdfkeywords={}, 
	pdfnewwindow=true,      
	colorlinks=true,       
	linkcolor=cyan,          
	citecolor=blue,        
	filecolor=pink,      
	urlcolor=cyan           
}
\usepackage[utf8]{inputenc} 
\usepackage[T1]{fontenc} 
\usepackage{ctable} 
\usepackage{pifont}

\usepackage{array}
\newcolumntype{L}{>{\centering\arraybackslash}m{3cm}} 
\usepackage{color, colortbl, xcolor, comment}
\usepackage{subfig}
\usepackage{threeparttable}
\usepackage{tcolorbox} 
\usepackage{algorithm}
\usepackage{chngcntr} 
\usepackage[noend]{algpseudocode}
\algrenewcommand\textproc{}
\algdef{SE}[SUBALG]{Indent}{EndIndent}{}{\algorithmicend\ }%
\algtext*{Indent}
\algtext*{EndIndent}
\usepackage{pdflscape}
\newtheorem{theorem}{Theorem}

\newtheorem{assumption}{Assumption}
\newtheorem*{corollary}{Corollary}

\newtheorem{remark}{Remark}
\newcommand{\xp}{x^{\psi}}
\newcommand{\xb}{x^{\beta}}
\newcommand{\ora}{\widehat\bpsi_{ora}}
\newcommand{\tru}{\bpsi^*}

\newcommand{\bx}{\textbf{\emph{x}}}
\newcommand{\by}{\textbf{\emph{y}}}
\newcommand{\bX}{\boldsymbol{X}}

\newcommand{\bY}{\textbf{\emph{Y}}}

\newcommand{\indep}{\perp \!\!\! \perp}
\newcommand{\ini}{\widehat\bpsi_{ini}}

\newcommand{\bI}{\boldsymbol I}
\newcommand{\bJ}{\boldsymbol J}

\newcommand{\bpsi}{\boldsymbol{\psi}}

\newcommand{\btau}{\boldsymbol{\tau}}

\newcommand{\bU}{\boldsymbol{U}}

\definecolor{Gray}{gray}{0.9}

\newcommand{\R}{{\mathbb R}}

\newcommand{\bbeta}{\boldsymbol{\beta}}

\newcommand{\expit}{\mbox{expit}}

\newcommand{\btheta}{\boldsymbol{\theta}}

\DeclareMathOperator*{\argmin}{arg\,min}
\DeclareMathOperator*{\argmax}{arg\,max}

\DeclarePairedDelimiter\abs{\lvert}{\rvert}%
\DeclarePairedDelimiter\norm{\lVert}{\rVert}%

\newcommand{\E}{\mathbb{E}}

\usepackage{lastpage}
\usepackage{fancyhdr}
\usepackage[parfill]{parskip} 
\usepackage[margin=1in]{geometry}
\usepackage{setspace}
\begin{document}
\def\spacingset#1{\renewcommand{\baselinestretch}%
{#1}\small\normalsize} \spacingset{1.4}
\date{}
\title{\bf Variable Selection for Individualized Treatment Rules with Discrete Outcomes}

\author{Zeyu Bian$^{1,6}$\footnote{Correspondence to: zeyu.bian@miami.edu}, 
Erica EM Moodie$^1$, Susan M Shortreed$^{2, 3}$, Sylvie D Lambert$^{4,5}$ and Sahir Bhatnagar$^{1}$ \\
$^1$Department of Epidemiology and Biostatistics, McGill University, Canada\\ $^2$Kaiser Permanente Washington Health Research Institute, USA \\ $^3$Department of Biostatistics, University of Washington, USA \\ $^4$Ingram School of Nursing, McGill University, Canada\\ $^5$St. Mary's Research Centre, Canada\\ $^6$Miami Herbert Business School, University of Miami, USA}


  \maketitle

\medskip
\begin{abstract}
An individualized treatment rule (ITR) is a decision rule that aims to improve individuals’ health outcomes by recommending treatments according to subject-specific information. In observational studies, collected data may contain many variables that are irrelevant to treatment decisions. Including all variables in an ITR could yield low efficiency and a complicated treatment rule that is difficult to implement. Thus, selecting variables to improve the treatment rule is crucial. We propose a doubly-robust variable selection method for ITRs, and show that it compares favorably with competing approaches. We illustrate the proposed method on data from an adaptive, web-based stress management tool.

\end{abstract}

\noindent%
\vfill

\newpage
\spacingset{2}
\section{Introduction}
In the precision medicine paradigm, treatment decisions are tailored to individuals rather than relying on a “one-size-fits-all” approach. This approach to treatment is beneficial when treatment effects are heterogeneous. 
For example, effective management of stress requires the development of personalized approaches, as patients with different characteristics respond to and engage with treatments differently.
With the aim of improving individuals’ health outcomes, individualized treatment rules (ITRs) \citep{Murphy,gest,DTRs,kosorok2015adaptive} recommend effective treatments based on each person's specific characteristics. However, collected data often contain many variables that are irrelevant for tailoring treatment. Including all variables in an analysis could reduce statistical efficiency by estimating unnecessary coefficients whose estimates fluctuate around zero for variables that are not useful for tailoring treatment, and yielding an unnecessarily complicated treatment decision rule that is difficult for physicians to interpret or implement. It is therefore important to develop variable selection methods with the objective of optimizing individuals' outcomes by identifying useful tailoring variables.  

Variable selection for ITRs has been studied in \citet{lassoq2,jeng2018high,al,bian}, all of which focus on penalized regression-based estimation methods. \citet{jeng2018high} and \citet{lassoq2} considered only a singly robust method in which the propensity score must be correctly specified. \citet{al} used the Dantzig selector directly to penalize the A-learning \citep{gest} estimating equation; \citet{bian} used penalized dynamic weighted ordinary least squares regression to perform variable selection. \citet{zhang2018variable} and \citet{zhang2022subgroup} extended the classification framework for estimating optimal treatment regimes in \citet{zhang2012estimating} to a setting in which variable selection can be performed. In \citet{zhang2018variable}, variables are sequentially selected based on the additional improvement provided by the new variable; while \citet{zhang2022subgroup} added a penalized term for the objective function to select the important variables. The methods considered in \citet{zhang2018variable, al, zhang2022subgroup, bian} are all doubly robust, i.e., they yield consistent estimators while requiring only one of two nuisance models to be correct. 

All of the afore-mentioned methods focus solely on the case in which the outcome is continuous. Discrete outcomes introduce additional computational challenges to the estimation of ITRs and the variable selection procedure, due to the common use of a non-identity link function. Existing literature focusing on discrete outcomes ITR estimation includes Q-learning \citep{DTRs,linn2017interactive}, Bayesian additive regression trees \citep{logan2019decision}, and A-learning \citep{robins1992estimating, tchetgen2010doubly}. However, none of these approaches has been extended to include variable selection. \citet{tian2014simple} proposed a straightforward method for estimating ITRs while performing tailoring variable selection by omitting all main effect terms for covariates, and re-scaling the covariates in the interaction terms. In this approach, the binary outcome case was also considered, although only in the randomized-treatment setting. \citet{chen2017general} further generalized the method in \citet{tian2014simple} to observational studies for ITR estimation and variable selection. Nevertheless, the proposed approach for binary outcomes requires the propensity score to be correctly specified. A further augmentation of \citet{tian2014simple} and \citet{chen2017general} was discussed in \citet{chen2017general} for binary outcomes, yet this augmentation approach still cannot achieve the desired double robustness property, because of the use of the nonlinear loss function (see Remark 2 in \citet{tian2014simple} for a more detailed explanation). In other words, the augmentation idea for binary outcomes in \citet{chen2017general} is used mainly for the resulting efficiency gain; a correct specification of the treatment model is still needed for consistent ITR estimation even if the outcome model is correctly specified. In this article, we focus on developing doubly robust ITR estimation with variable selection for discrete outcomes (count and binary outcomes).

To provide robustness against model misspecification, ITRs are often estimated using estimating equations \citep{Murphy,gest}. There are at least two ways to achieve sparsity in the use of estimating equations: via a Dantzig selector \citep{dan} or by a regularized estimating equation (REE). Denote by $\bU(\btheta)\in \mathbb{R}^p$ an estimating equation, where $\btheta \in \mathbb{R}^p$. The Dantzig estimator $\widehat\btheta_{dan}$ can be found by solving the constrained optimization problem: $\widehat\btheta_{dan}=\argmin_{\btheta}\: \norm{\btheta}_1$, subject to $\norm{\bU(\btheta)}_{\infty} \leq n\lambda$, where $\lambda$ is a tuning parameter used to control sparsity, and $n$ is the sample size. Another way to induce sparsity is to solve the REE: $\bU(\btheta)=n\lambda q (|\btheta|),$ where $q(|\cdot|)$ is the subgradient of a penalty function $\rho (|\cdot|)$, i.e., $q(|\cdot|) = \partial \rho (|\cdot|)$. For example, lasso \citep{lasso} regression defined by $\min_{\bbeta} \{\lVert \bY-\bX\btheta \rVert_2^2+n\lambda\lVert \btheta\rVert_1\}$ is a special case of the REE $\bU(\btheta)=n\lambda\partial \norm\btheta_1$, where $\bU(\btheta)=\bX^T(\bY-\bX\btheta)$, $\bX \in \mathbb{R}^{n \times p}$ is the design matrix, and $\bY\in \mathbb{R}^n$ is the response.

While the Dantzig selector and REE work well for continuous outcomes \citep{al}, their implementation in ITRs can be difficult for discrete outcomes, which are usually modeled with nonidentity link functions. Indeed, the existing doubly robust estimating equations to estimate ITRs for discrete outcomes are nonlinear \citep[see later in Section \ref{dis}]{robins1992estimating, tchetgen2010doubly}, and hence the Dantzig selector cannot be solved using linear programming \citep{james2009generalized}. As for REE, it has been studied in \citet{johnson2008penalized} and \citet{wang2012penalized} using local quadratic approximation \citep{SCAD} to solve the REE, which is computationally burdensome since it requires the calculation of the inverse of the Hessian matrix. Finally, even if the solution of the Dantzig selector or the REE can be found, selecting the tuning parameter in an ITR context is challenging since the goal is inference about treatment effects rather than just predictive performance. This means that we cannot simply select the tuning parameter that has the lowest prediction error as in the more classical prediction setting.

Our work proposing new doubly robust estimating functions for count and binary outcomes is motivated by the desire to evaluate the effectiveness of a web-based stress management intervention for individuals with cardiovascular disease. We use longitudinal data collected as part of a two-stage pilot sequential multiple assignment randomized trial \citep{lambert2021adaptive} for estimating a stress management ITR. Due to the small sample size of the study (50 observations) and relatively large number of potentially relevant variables collected, 
selecting useful variables for tailoring treatment solely based on expert knowledge can be an extremely challenging task. Our newly proposed estimating equations allow integration of variable selection approaches. We apply this variable selection approach with our proposed algorithm for solving the proposed estimating equations to provide valuable insights into the influence of various tailoring variables on patient outcomes, enabling the development of more effective and personalized approaches to the stepped-care approach for web-based stress management.

Specifically, in this work, we propose two new, doubly robust estimating functions for count and binary outcomes respectively, in the setting of binary treatment with a single stage, to estimate an ITR. A benefit of our proposed estimating function is that it can be easily generalized to a penalized framework, which permits estimating the optimal treatment regimes and selecting important tailoring variables simultaneously. We show that with a suitable choice of weights, a simple penalized regression model for estimating an ITR enjoys the desired double robustness property and is straightforward to implement. The advantage of the newly proposed approach compared to alternative regularized ITR estimation methods is that it can be viewed from a minimization perspective. Hence, the implementation is simple, various penalty functions can be used, and the solution can be found using existing computationally efficient tools in standard software. We propose a tuning parameter selection procedure to address that the goal of an ITR analysis is estimating a decision rule rather than prediction. To our knowledge, doubly robust variable selection in ITR estimation for discrete outcomes has not been studied in existing literature. 

The rest of this article is organized as follows. In Section 2, we present introductory concepts and review existing doubly robust estimation methods for discrete outcomes. In Section 3, we introduce our proposed estimation methods, and we extend them to a penalized framework in Section 4, followed by statements of theoretical properties. A number of simulation studies are in Section 5. Finally, in Section 6, we apply our method to data from an adaptive web-based stress management study.

\section{Background}
\subsection{Notations, Assumptions and Introductory Concepts}

Throughout, we use uppercase letters to denote random variables and lowercase letters to denote observed values. We use nonbold letters to denote individual-level data and bold letters to denote all observations in the data, e.g., $X_i \in \mathbb{R}^p$ are the covariates for subject $i$, while $\bX \in \mathbb{R}^{n\times p}$ are covariates for all subjects. In a single stage ITR, $V_i=(X_i, A_i, Y_i)$ consists of the data for the $i$th subject, where $X_i$ is the subject’s baseline covariates, $A_i$ is the binary treatment received, and $Y_i$ is the subject’s outcome. Throughout, we consider binary treatment in a static setting (single stage ITR), while extension to general discrete allocations is discussed in Section \ref{sec: discussion}. In the sequel, we will suppress subscript $i$ where it is clear. We denote the potential outcome under the treatment $a$ as $Y^a$. The objective of an ITR analysis is to find the optimal treatment $d^{opt}(X)$ such that the expected potential outcome $\E(Y^{d})$ is maximized across the population of individuals. To estimate ITRs, we assume the following standard causal assumptions: (1) the stable unit treatment value assumption (SUTVA) \citep{rubin1980}: an individual’s potential outcome is not affected by other subjects’ treatment assignments; (2) consistency: $Y=AY^1+(1-A)Y^0$; (3) conditional exchangeability \citep{ignore}: $Y^a \indep A |X=x$; and (4) positivity: $P(A=a|X=x)>0$ almost surely for all $x$ and $a=0,\,1$.

Finally, we assume that the observations $V_i, \; i=1, \dots, n$ are independent and identically distributed with probability density $h(V)$ with respect to a measure $\nu$. Moreover, we assume the relationship between $Y$ and $(X, A)$ can be captured by a semiparametric regression model: $g\big(\E(Y^a|X=x)\big)=g\big(\E(Y|X=x,A=a)\big)=f_0(x;\bbeta)+\gamma\left(x, a; \boldsymbol{\psi}\right)$, where $g$ is a known link function, $f_0$ is an unknown baseline function, and $\gamma$ is a known function that satisfies $\gamma\left(x, 0; \boldsymbol{\psi}\right)=0$, which is referred to as the blip function \citep{gest}. A blip function can be interpreted as the difference on the linear predictor scale of the transformed mean potential outcomes \begin{align*}
    \gamma\left(x, a\right)&=g\big(\E(Y^a|X=x)\big)-g\big(\E(Y^0|X=x)\big)\\
   &=g\big(\E(Y^a|X=x,A=a)\big)-g\big(\E(Y^0|X=x,A=0)\big).
\end{align*} In this modeling paradigm, $f_0$ is irrelevant for making treatment decisions (a nuisance model). Hence, our parameter of interest is $\bpsi$, and the optimal ITR $d^{opt}(x)$ is given by \begin{align*}
    d^{opt}(x)=\argmax_d \E(Y^d)=\argmax_d \E_X\left\{\E\left[g^{-1}(f_0(X;\bbeta)+\gamma\left(X, d(X); \boldsymbol{\psi}\right))\right]|X\right\}\\= \argmax_d \E_X\left[ \, f_0(X;\bbeta)+\gamma\left(X, d(X); \boldsymbol{\psi}\right)\right]=\argmax_d \gamma\left(X, d(X); \boldsymbol{\psi}\right)= \mathbb{I}(\gamma\left(X, 1; \boldsymbol{\psi}\right)>0),
\end{align*}
given an increasing link function. Throughout, we assume a log link for count outcomes and a logit link for binary outcomes.

\subsection{Existing Estimation Methods for Discrete Outcomes}\label{dis}

\subsubsection{A-learning for Count Outcomes}

Denote by $\xp$ the covariates in the blip model and by $\xb$ the covariates in the baseline model; in what follows, the superscript is omitted if they are identical. We assume that the blip function is of the form of $\gamma(\xp,a;\bpsi)=a\bpsi^T \xp$ in the sequel. Then the A-learning estimating equation \citep{robins1992estimating} for a count outcome, with a log link function, is
\begin{gather*}
    \bU_1(\bpsi)=\frac{1}{n}\sum_{i=1}^n x^{\psi}_i  (a_i- \widehat\pi_i)\exp\{-\gamma(\xp_i,a_i;\bpsi)\}\left(y_i-\exp(f(\xb_i;\widehat \bbeta)+\gamma(\xp_i,a_i;\bpsi))\right), 
\end{gather*} where $f$ is the posited baseline model (not necessarily identical to $f_0$), $\widehat\bbeta$ is a plug-in estimator, and $\widehat\pi$ is the estimated propensity score. The propensity score \citep{ps} is defined as the coarsest balancing score $b(x)$ such that $b(x)=P(A=1|x)$, i.e., the probability of treatment received conditional on confounders. In observational studies, this quantity is unknown and needs to be estimated from the data. It can be shown that $\bU_1(\bpsi)$ is an unbiased estimating equation \citep{robins1992estimating}, provided that at least one nuisance model (propensity score model or baseline model) is correctly specified. This property is the so-called double robustness property \citep{bang2005doubly}. Since in observational studies, one can never be sure that
either a baseline model or a propensity score model is correct, a double robustness estimator hence is highly desirable, as it provides some safeguards against model mis-specification. Furthermore, in settings such as our motivating example, where treatment is randomized, doubly robust methods ensure consistency since the treatment allocation model is known by design.

\subsubsection{A-learning for Binary Outcomes}

Estimation is more complicated when the outcome is binary; the blip parameter is estimated by solving the following estimating equation, assuming a logit link function: $$\bU_2(\bpsi)=\frac{1}{n}\sum_{i=1}^n x^{\psi}_i (a_i- \widehat\pi^*)\left(y_i-\mbox{expit}(f(\xb_i;\widehat \bbeta)+\gamma(\xp_i,a_i;\bpsi))\right),$$ where $$\widehat\pi^*=\left(1+\frac{(1-\mbox{expit}(u(x;\widehat \btau)) \expit(f(x;\widehat \bbeta))}{\expit(u(x;\widehat \btau))\expit(f(x;\widehat \bbeta)+\gamma(x,a;\bpsi))}\right)^{-1},$$ $\mbox{expit}(t)=\frac{\exp(t)}{1+\exp(t)}$, and $u(x;\btau)$ is the nuisance treatment model of $\mathbb{E}(A|Y=0,X)$. \citet{tchetgen2010doubly} showed that $\bU_2(\bpsi)$ is an unbiased estimating equation when at least one of $\mathbb{E}(Y|X,A=0)$ or $\mathbb{E}(A|X,Y=0)$ is correctly specified. Note that for the logit link, the quantity $\E(A|Y=0,X)$ is modeled instead of the propensity score to assure the double robustness property, because of the symmetry property of the odds ratio: $$e^{X^\top\bpsi}=\frac{P(Y=1|A=1,X)P(Y=0|A=0,X)}{P(Y=0|A=1,X)P(Y=1|A=0,X)}=\frac{P(A=1|Y=1,X)P(A=0|Y=0,X)}{P(A=0|Y=1,X)P(A=1|Y=0,X)}.$$ \citet{chen2007semiparametric} showed that there are at least two ways to study the association parameter (in our case, the blip parameter): through the density of $Y$ given $X$ and $A$ or through the density of $A$ given $X$ and $Y$. This provides an intuitive explanation of why $\mathbb{E}(Y|X,A=0)$ and $\mathbb{E}(A|X,Y=0)$ are modeled to assure the double robustness property. 

As noted above, the implementation of the Dantzig selector or the REE can be difficult for the A-learning estimating function. In the next section, we propose an alternative estimation method that is also doubly robust and can easily accommodate variable selection. 

\section{Doubly Robust Weighted Generalized Linear Model}

In this section, we propose two new estimating equations for count and binary outcomes, respectively, and we show that solving these two estimating equations can be reformulated as an iteratively reweighted generalized linear model (IRGLM). The obtained estimators are doubly robust, and the proposed estimating equation can be easily generalized to a variable selection framework. Throughout, we posit a linear model for the baseline function, i.e., $f(x;\bbeta)=x^T\bbeta$, which is not necessarily identical to the true baseline model $f_0$.

\subsection{Count Outcomes}
For count outcomes, we present the following estimating function:
 \begin{gather*} 
    \bU_3(\bbeta,\bpsi)=\sum_{i=1}^n \begin{pmatrix}a_ix^{\psi}_i\\ \xb_i \end{pmatrix}  |a_i- \widehat\pi_i| \exp\{-\gamma(\xp_i,a;\bpsi)\}\left(y_i-\exp(f(\xb_i; \bbeta)+\gamma(\xp_i,a;\bpsi))\right).
\end{gather*} This estimating equation is inspired by the A-learning estimating equation $\bU_1(\bpsi)$ and the weighted least squares equation using overlap weights $|a_i-\pi_i|$ in \citet{wallace2015doubly}. The overlap weights $|a_i-\pi_i|$ ensure that the above estimating equation is unbiased even if the baseline model is misspecified (under the setting that $\pi$ is correctly specified). Moreover, \citet{wallace2015doubly} empirically demonstrated that the use of overlap weights can improves efficiency of the resulting estimator over estimators of the same form that use alternative weights such as inverse probability of treatment weights. Note that this equation takes a similar form to $\bU_1(\bpsi)$, with the leading term $\exp\{-\gamma(\xp_i,a;\bpsi)\}$, and shares a similar form to \citet{wallace2015doubly} using overlap weights, but is not identical to either. 

\begin{assumption}
When at least one of the two nuisance models $\pi$ or $f$ is correctly specified, there exists a unique population parameter $\btheta^*=(\bbeta^*,\bpsi^*)$ such that $\E[\bU_3(\bbeta^*,\bpsi^*)]=\boldsymbol0$.
\end{assumption}

\begin{theorem}
Assume that the SUTVA, ignorability, consistency, and positivity conditions described in Section 2.1 and \textbf{Assumption 1} hold as described in Section 3.1. If the posited baseline model satisfies $x^\psi \subseteq x^\beta$, and the link function $g$ is known, then the solution $\bpsi^*$ to $\E[\bU_3(\bbeta,\bpsi)]=0$ satisfies $\bpsi^*=\bpsi_0$, where $\bpsi_0$ is the underlying true blip parameter.
\end{theorem}

Theorem 1 states that under standard causal assumptions, the population parameter $\bpsi^*$ is equivalent to the true data-generating value of the blip (and corresponding ITR) parameter $\bpsi_0$, if one of two nuisance models, $\pi$ or $f$, is correctly specified. This implies that the blip estimator $\widehat\bpsi$ obtained by solving $\bU_3(\bbeta,\bpsi)$ is a doubly robust estimator.

\begin{remark}
The condition of the existence of a unique population parameter is similar to the condition of the existence of the quasi-maximum likelihood estimate when the likelihood is misspecified \citep{white1982maximum}. The assumption that $\xp \subseteq \xb$ in the posited model is referred to as the strong heredity assumption \citep{strong}: the corresponding main effects of an interaction term must be included in the model.
\end{remark}

Now we demonstrate that $\bU_3(\bbeta,\bpsi)$ can be specified as an IRGLM for which efficient computational solutions exist, and thus a penalized estimator can be constructed from the penalized generalized weighted linear model accordingly. We propose Algorithm \ref{alg:alg1} to solve $\bU_3(\bbeta,\bpsi)$. The key is to treat the $|a_i- \widehat\pi_i| \exp\{-\gamma(\xp_i,a;\bpsi)\}$ term in $\bU_3(\bbeta,\bpsi)$ as a constant in each iteration $t$. In this way, Step $7$ in Algorithm \ref{alg:alg1} is equivalent to a weighted generalized linear model (GLM) with weights $|a_i- \widehat\pi_i| \exp\{-\gamma(\xp_i,a;\widetilde\bpsi_{t})\}$, where $\widehat\pi$ is the estimated propensity score that does not change across iterations and $\widetilde\bpsi$ is the current value of the blip parameter estimate from the most recent iteration update. This can be solved efficiently using, for example, the \texttt{glm} function in $\textbf{\textsf{R}}$ and specifying the \texttt{weights} argument. 

\begin{algorithm}[t]
	\caption{}
	\begin{algorithmic}[1]
		\small
		\Function{}{$ x_i,a_i,y_i,\widehat\pi_i,\varepsilon$}
        \State Set iteration counter $t \gets 0$
		\State Initialize $\widetilde\bpsi_0$
		\State  $w_{i0} \gets |a_i- \widehat\pi_i| \exp\{-\gamma(\xp_i,a_i;\widetilde\bpsi_0)\}$ for $i=1, \ldots, n$
		\Repeat
		\State Solve $\bbeta_t$ and $\bpsi_t$ such that 
		\State $\sum_{i=1}^n \begin{pmatrix}a_ix^{\bpsi}_i\\ \xb_i \end{pmatrix}  w_{it}\left(y_i-\exp(f(\xb_i; \bbeta_t)+\gamma(\xp_i,a_i;\bpsi_t))\right)=0$
		\State $\widetilde\bpsi_{t+1} \gets \bpsi_t$
		\State $w_{i(t+1)} \gets |a_i- \widehat\pi_i| \exp\{-\gamma(\xp_i,a_i;\widetilde\bpsi_{t+1})\}$ 
		\State $t \gets t + 1$ \Until{$\norm{\bpsi_t-\bpsi_{t-1}}<\varepsilon$}
		\EndFunction
\end{algorithmic}
\label{alg:alg1}
\end{algorithm}

\subsection{Binary Outcomes}
A similar framework can be built for binary outcomes using the logit link function. We present estimating equation $\bU_4(\bbeta,\bpsi)$ for binary outcomes:
 \begin{gather*} 
    \bU_4(\bbeta,\bpsi)=\sum_{i=1}^n \begin{pmatrix}a_ix^{\psi}_i\\ \xb_i \end{pmatrix}  |a_i- \widehat\pi_i^*| \left(y_i-\expit(f(\xb_i; \bbeta)+\gamma(\xp_i,a;\bpsi))\right),
\end{gather*} where $$\widehat\pi^*=\left(1+\frac{(1-\mbox{expit}(u(x;\widehat \xi)) \expit(f(x; \widehat\bbeta^*))}{\expit(u(x;\widehat \xi)\expit(f(x;\widehat\bbeta^*)+\gamma(x,1;\bpsi))}\right)^{-1},$$ and $u(x;\xi)$ is the nuisance treatment model for $\mathbb{E}(A|Y=0,X)$. Under mild conditions, the solution of $\bU_4(\bbeta,\bpsi)$ is a doubly robust estimator. Note that all theoretical properties for count outcomes can be applied equally to binary outcomes; for convenience and space, we include the results for binary outcomes in the Appendix (Section A). Algorithm \ref{alg:alg2} can be used to solve $\bU_4(\bbeta,\bpsi)$, once again treating the term $|a_i-\widehat\pi_i^*|$ as a constant in each iteration.

\begin{algorithm}[t]
	\caption{}
	\begin{algorithmic}[1]
		\small
		\Function{}{$ x_i,a_i,y_i,\widehat\pi_i,\varepsilon$}
        \State Set iteration counter $t \gets 0$
		\State Initialize: $\widetilde\bpsi_0$
		\State  $w_{i0} \gets |a_i- \widehat\pi_i^*(\widetilde\bpsi_0)|$ for $i=1, \ldots, n$
		\Repeat
		\State Solve $\bbeta_t$ and $\bpsi_t$ such that 
		\State $\sum_{i=1}^n \begin{pmatrix}a_ix^{\psi}_i\\ \xb_i \end{pmatrix}  w_{it}\left(y_i-\expit(f(\xb_i; \bbeta_t)+\gamma(\xp_i,a_i;\bpsi_t))\right)=0$
		\State $\widetilde\bpsi_{t+1} \gets \bpsi_t$
		\State $w_{i(t+1)} \gets |a_i- \widehat\pi_i^*(\widetilde\bpsi_{t+1})|$ 
		\State $t \gets t + 1$
		\Until{$\norm{\bpsi_t-\bpsi_{t-1}}<\varepsilon$}
		\EndFunction
\end{algorithmic}
\label{alg:alg2}
\end{algorithm}

\section{Tailoring Variable Selection}

In this section, we introduce sparsity to our proposed estimating function using the formulation of a REE, and show that this REE is asymptotically equivalent to a penalized weighted GLM given an appropriate initial estimator. Throughout, the main effect of the treatment $A$ is not penalized, as our goal is to select the important tailoring variables.

\subsection{Penalized Doubly Robust Method}

Due to the nonlinear part (log or logit link) of the estimating equation for discrete outcomes, a Dantzig selector with A-learning estimating equation $\bU_1(\bpsi)$ or $\bU_2(\bpsi)$ cannot be solved using linear programming \citep{james2009generalized}. Hence, we pursue an REE approach to introduce sparsity to the proposed estimating equations $\bU_3(\bbeta,\bpsi)$ and $\bU_4(\bbeta,\bpsi)$, and once again, reformulate the REE as a penalized weighted GLM. We call this approach the penalized doubly robust (PDR) method, as it will be shown later that the penalized estimator obtained by solving the ITR REE is a doubly robust estimator.

For count and binary outcomes, ITR REE requires finding the solution of, respectively, \begin{gather}
    \sum_{i=1} \begin{pmatrix}a_ix^{\psi}_i\\ \xb_i \end{pmatrix}  |a_i- \widehat\pi_i| \exp\{-\gamma(\xp_i,a;\bpsi)\}\left(y_i-\exp(f(\xb_i; \bbeta)+\gamma(\xp_i,a;\bpsi))\right)=n\lambda q (|\btheta|), \label{eq:ree}
\end{gather} and \begin{gather}
    \sum_{i=1} \begin{pmatrix}a_ix^{\psi}_i\\ \xb_i \end{pmatrix}  |a_i- \widehat\pi_i^*| \left(y_i-\expit(f(\xb_i; \bbeta)+\gamma(\xp_i,a;\bpsi))\right)=n\lambda q (|\btheta|). \label{eq:ree2}
\end{gather} 



To estimate the blip parameters, $\bpsi$, consistently, we require that the penalized model satisfies the following properties: (a) no false exclusion of tailoring variables, and (b) the selected model has the strong heredity property, i.e., $\widehat\psi_j \neq 0 \implies \widehat\beta_j\neq 0$ (i.e., without loss of generality, assume that $x^{\psi}$ has the same ``ordering'' as $x^{\beta}$). Many penalty functions can yield a model that has variable selection consistency, i.e., no false inclusion and no false exclusion; for example, lasso, SCAD \citep{SCAD}, and adaptive lasso \citep{adaptive}. However, these methods all fail to achieve the strong heredity property. Thus, further work is required to implement them in this setting. \citet{bian} used reparametrization to ensure strong heredity when using penalization in the context of ITR. Here, we modify the adaptive lasso penalty and show using these modified adaptive weights allows not only the strong heredity constraint to be met, but also the (asymptotically) unbiased estimation  of blip parameters.

We omit the subscript for the estimating functions $\bU_3(\bbeta,\bpsi)$ and $\bU_4(\bbeta,\bpsi)$ for now, as the properties for both count and binary outcomes can be developed using a general notation $\bU(\bbeta,\bpsi)$. Let $\btheta_0=(\bbeta_0, \bpsi_0)$ denote the underlying true parameters and recall that $\btheta^*=(\bbeta^*, \bpsi^*)$ is the unique population parameter such that $\E[\bU(\bbeta^*,\bpsi^*)]=0$. Let $s$ be the number of nonzero components of $\bpsi_0$ (or equivalently, $\bpsi^*$) and $S$ denote the set of indices of nonzero components for $\bpsi_0$. Denote by $S'$ the set of indices of nonzero components for $\bbeta^*$. To satisfy the strong heredity property, we want the estimated baseline model to satisfy $\widehat \bbeta_{\widetilde S}\neq 0$ as $n$ goes to infinity, where $\widetilde S=S\cup S'$ (as such, $S \subseteq \widetilde S$ and hence strong heredity holds). The goal is to estimate a targeted indices set $S^*$, such that $\widehat\btheta_{S^*}\neq 0$ and $\widehat\btheta_{S^*_c}= 0$ with probability tending to $1$, where $S^*_c$ is the complement of $S^*$ (note that $\btheta^*_{S^*}=(\bbeta^*_{\widetilde S},\bpsi^*_S)$). 


Suppose we have an initial estimator $\widehat\btheta_{ini}=(\widehat\bbeta_{ini}, \ini)$, such that $\sqrt{n} \norm {\widehat\bbeta_{ini}-\bbeta^*}=O_p(1)$ and $\sqrt{n}\norm {\ini-\tru}=O_p(1)$. Following the adaptive lasso \citep{adaptive} principle, we construct our adaptive weights for the corresponding coefficients $\bbeta$ and $\bpsi$ as follows: \begin{gather}
    \widehat \omega_j^{\beta}=\left\{\max \left(|\widehat\beta_j^{ini}|,|\widehat\psi_j^{ini}|\right)\right\}^{-1} \mbox{ and  } \widehat \omega_j^{\psi}=\abs{\widehat\psi_j^{ini}}^{-1}  \label{eq:mod}.
\end{gather}We then use the penalty function $\rho(|\btheta|)=\rho(|\bbeta|)+\rho(|\bpsi|)$, where \begin{gather*}  \rho(|\bbeta|)=\sum_{j=1}^p \widehat \omega_j^{\beta}|\beta_j|  \mbox{ and  }  \rho(|\bpsi|)=\sum_{j=1}^p \widehat \omega_j^{\psi} |\psi_j|.\end{gather*}In this way, for nonzero coefficients of blip variables, the associated weights and those of their corresponding main effects both converge to finite constants, and thus always remain in the model. We refer to our proposed weights in Expression \eqref{eq:mod} as modified adaptive weights, since these build on the adaptive lasso framework but differ in the choice of $\widehat \omega_j^{\beta}$. Theorem 2 establishes the existence of a $\sqrt{n}$-consistent solution to the ITR REE \eqref{eq:ree} and \eqref{eq:ree2}.


\begin{theorem}[\textit{\textbf{Existence and Selection Consistency}}]
Assume that conditions in Theorem 1 hold, penalty functions are constructed using the modified adaptive weights described in Expression \eqref{eq:mod}, and the tuning parameter satisfies $\sqrt{n}\lambda=o(1)$ and $n\lambda\to \infty$. There then exists a $\sqrt{n}-$consistent solution $\widehat \btheta=(\widehat \bbeta,\widehat \bpsi)$ of the ITR REE, such that $\widehat \bpsi_S \neq 0$ and $\widehat \bpsi_{S_c}=0$.
\end{theorem}

By Lemma 1 in the Appendix B.2, to establish the existence of the REE solution, it suffices to show that for sufficiently large $n$, there exists a constant $r$, such that on the boundary of a ball around $\btheta^*$ with radius $n^{-1/2}r$, the variational inequality holds for function $\bU(\btheta)-n\lambda q(|\btheta|)$ with high probability. That is, for any $\varepsilon>0$, $$\mathbb{P}\left(\inf_{\lVert \btheta-\btheta^* \rVert=n^{-1/2}r}(\btheta-\btheta^*)^T [\bU(\btheta)-n\lambda q(|\btheta|)] >0\right) > 1-\varepsilon.$$ This technique has been adopted in \citet{portnoy1984asymptotic} and \citet{wang2011gee} to prove the existence of the $M$-estimator and generalized estimated equations estimator when the number of predictors is large. Theorem 3 establishes the asymptotic normality of the ITR REE estimators under standard regularity conditions (see Appendix B for details).

\begin{theorem}[\textit{\textbf{Asymptotic Normality}}] 
For any $\sqrt{n}-$consistent solution $\widehat \btheta$ of ITR REE,$$\sqrt n \bJ(\bpsi^*_S)\{\widehat \bpsi_S-\bpsi^*_S+\bJ(\bpsi^*_S)^{-1}\lambda \,q(|\bpsi^*_S|)\}\to_d N\big( 0,\bI(\bpsi^*_S)\big),$$ where $\bI(\btheta)\in \R^{2p\times 2p}$ is the variance of the estimating equation $\bU(V_i,\btheta)$, $\bJ(\btheta)\in \R^{2p\times 2p}$ is the quantity $\E_{\btheta}\left[-\frac{\partial \bU(V_i,\btheta)}{\partial \btheta }\right]$, $p$ is the length of the full covariate vector $X_i$, and $\bI(\bpsi^*_S)$ and $\bJ(\bpsi^*_S)$ are the corresponding $s\times s$ sub-matrices of $\bI$ and $\bJ$ evaluated at the truth.
\end{theorem}

A detailed proof of Theorem 2 and Theorem 3 are in the Appendix (Sections B.4 and B.5). To illustrate the double robustness property of our proposed estimators, we borrow the idea of the oracle estimator \citep{SCAD}. Define the oracle estimator  $\ora \in \R^s$ as the solution of $\bU(\bbeta,\bpsi)$ using $f(x_{\widetilde S})$ and $\gamma(x_{S}, a)$ (i.e., assume that the zero and nonzero coefficients are known in advance). Since we do not know the truly important variables in the application, the oracle estimator is just a concept to help establish the theoretical properties in variable selection. Due to the double robustness of $\bU(\bbeta,\bpsi)$, $\ora$ is a consistent asymptotically normal estimator of $\bpsi_{S}^*$ under standard regularity conditions for $M$-estimators. The properties of $\widehat \bpsi$ in Theorems 2 and 3 are referred to as the oracle property \citep{SCAD}, i.e., $\widehat \bpsi$ performs as well as the oracle estimator $\ora$.

\begin{corollary}[\textit{\textbf{Double Robustness}}]
The oracle estimator $\ora$ constructed above is a doubly robust estimator of $\bpsi_0$. Since the resulting estimator $\widehat \bpsi$ mimics the oracle estimator $\ora$, $\widehat \bpsi$ is also a doubly robust estimator. That is to say, the resulting estimator $\widehat \bpsi$ is a consistent estimator of $\bpsi_{0}$ if either of two nuisance models is correct. 
\end{corollary}

\subsection{A One-step Estimator}
\label{onestep}
For settings in which the number of variables, $p$, is fixed, we present an approximation to solve the ITR REE \eqref{eq:ree} in one step. Suppose that we can find an initial estimator $\ini$ of the blip parameter, such that $\sqrt{n}\norm {\ini-\tru}_2=O_p(1)$. Then we can plug $\ini$ into the weight term of Expression \eqref{eq:ree} and solve it directly, which is equivalent to maximizing a weighted penalized likelihood. Taking the count outcomes as an example, we can use the solution of the unpenalized estimating equation $\bU_1(\btheta)$ or $\bU_3(\bbeta,\btheta)$ as the initial estimator. Then under mild conditions, using $\ini$ as a plug-in estimator will have a negligible effect on the resulting estimator $\widehat\bpsi$. That is, the solution of $$
    \sum_{i=1} \begin{pmatrix}a_ix^{\psi}_i\\ \xb_i \end{pmatrix}  |a_i- \widehat\pi_i| \exp\{-\gamma(\xp_i,a_i;\ini)\}\left(y_i-\exp(f(\xb_i; \bbeta)+\gamma(\xp_i,a_i;\bpsi))\right)=n\lambda\partial \rho (|\btheta|)
$$ is asymptotically equivalent to the solution of \eqref{eq:ree}. In high dimensional settings in which an unpenalized initial estimator cannot easily be computed, the ridge penalty can be used to obtain the initial estimator.


\subsection{Tuning Parameter Selection} \label{tuning}

The choice of the tuning parameter $\lambda$ in Expressions \eqref{eq:ree} and \eqref{eq:ree2} plays an important role in the performance of the REE: An inappropriately large or small value of $\lambda$ will greatly weaken the performance of the resulting estimator in generating the estimation error and variable selection results. As previously noted, our proposed method can be viewed from a minimization perspective, i.e., $\widehat\btheta=\argmin_{\btheta} \{\mathcal{L}_n(\btheta;\by)+n\lambda \,\rho (|\btheta|)\}$. Following the idea used in classical information criteria \citep{akaike1974new,BIC}, we propose to select the tuning parameter by choosing the model that has the smallest value of $n^{-1}[D_\lambda(\widehat\btheta,\by)+\kappa_n s_\lambda]$, where $D_\lambda(\widehat\btheta,\by)=2[\mathcal{L}_n^{sat}(\widehat\btheta;\by)-\mathcal{L}_n(\widehat\btheta;\by)]$ is the quasi-deviance, $\mathcal{L}_n^{sat}$ is the quasi-log-likelihood of the saturated model, $\kappa_n$ is some positive sequence, and $s_\lambda$ is the number of nonzero components in the model for a given $\lambda$. We suggest setting $\kappa_n$ as $\log (\log n) \log p\;$following \citep{fan2013tuning}, as this can achieve model selection consistency in a penalized likelihood setting. In practice, we could also use cross-validation to choose the tuning parameter that corresponds to the lowest average loss $\mathcal{L}^{cv}_n(\widehat\btheta;\by)$.

In a penalized likelihood, where the goal is prediction, the optimal $\lambda$ is often chosen so the corresponding model has the lowest information criterion, usually estimated by a measure of model fit (e.g., negative log-likelihood) with an extra penalty term such as the Akaike information criterion \citep{akaike1974new} or the Bayesian information criterion \citep{BIC}. However, using an Akaike or Bayesian information criterion to select the tuning parameter will fail if the likelihood is misspecified (i.e., outcome model is misspecified). Thus, these classic methods of tuning parameter selection are not appropriate to the doubly robust setting where a likelihood is not positive and the mean model is not assumed to be correctly specified. Our proposed approach to selecting the tuning parameter outlined above requires that only one of the nuisance models is correctly specified.


\section{Numerical Studies}\label{sim3}

In this section, we illustrate the double robustness of our proposed method and show how the choice of the initial estimator can impact the resulting estimators. 

\textit{\textbf{Competing methods and implementation:}} we compare our proposed method with three different methods: unpenalized A-learning, \citet{zhang2018variable} and \citet{zhang2022subgroup}, where the last two competing methods were established based on the binary classification framework proposed in \citep{zhang2012estimating}. The $\textbf{\textsf{R}}$ package \texttt{drgee} \citep{zetterqvist2015doubly} is implemented to obtain the A-learning estimates; in addition, the sample code to conduct methods in \citet{zhang2018variable} and \citet{zhang2022subgroup} can be found in the supplementary material for the latter article.

Recall that in Section \ref{onestep}, the initial estimator can be obtained from A-learning or our proposed IRGLM. We now evaluate the performance of our proposed PDR method using two different initial estimators for the variable selection rate and the resulting error rate in the estimated treatment decision, as well as for the value function (expected outcome) of the estimated decision rules. The error rates and the average value function were calculated over a testing set of size 10,000. The data generation procedure for count outcomes is 
\begin{itemize}
\item Step 1: Generate $15$ independent multivariate normal covariates ($X_1,\dots,X_{15}$) with mean equal to 0.5 and unit variance. 
\item Step 2: Generate treatment such that $P(A=1|x_1,x_2)= \expit(-0.2+\sum_{j=1}^2 x_j)$. 
\item Step 3: Set the blip function as $\gamma(x,a;\bpsi)=a(\psi_0+ \psi_1x_1)$ for $\psi_0=1$ and $\psi_1=-2$. 
\item Step 4: Set the baseline model to $f(\bx;\bbeta)=\exp(-x_1^2-x_2^2+x_3-x_4)+x_1-0.2x_2$. 
\item Step 5: Generate the outcome $Y \sim  \mbox{Poisson}(\exp\left(f(\bx;\bbeta)+\gamma(\bx,a;\bpsi)\right)).$
\end{itemize} Under this data generation procedure, the optimal treatment is $\mathbb{I}(1-2x_1 > 0)$, which corresponds to treatment $A=1$ for about $50\%$ of subjects, and the marginal mean of the outcome under observed (rather than optimal) treatment is 1.21.

The data generation procedure for binary outcomes is the same for steps 1-3 above. In Step 4, we now set the nuisance treatment model as $\E(A|Y=0,X=x)=\exp(-x_1^2-x_2^2+x_3-x_4)+x_1-0.2x_2$, and marginalize the conditional expectation over the distribution of $Y$ to obtain the propensity score model $\E(A|X=x)$. Lastly, we generate the outcome $Y \sim  \mbox{Bernoulli}(\expit\left(f(\bx;\bbeta)+\gamma(\bx,a;\bpsi)\right)).$ Under this data generation procedure, the optimal treatment corresponds to treatment $A=1$ for about $50\%$ of subjects, and the marginal mean of the outcome under observed (rather than optimal) treatment is 0.47.

For both count outcomes and binary outcomes, we consider two scenarios with two sample sizes (500 and 1000). The baseline model is misspecified in scenario 1 (a linear working model is used), and the treatment model is misspecified in scenario 2 (the propensity score is setting to $0.5$ for all the observations). For PDR, we consider two alternative initial estimators: In the first case, referred to as PDR1, the estimator is obtained from A-learning, and in the second, PDR2, from our proposed IRGLM approach. Finally, we refer to unpenalized A-Learning and the methods in \citet{zhang2018variable} and \citet{zhang2022subgroup} as UA, ZZ1, and ZZ2, respectively.

\begin{table}[H]
\caption{Error rate (ER), value, false-negative (FN) and false-positive (FP) rate of variable selection results, with $n=500$ and 1000, for 400 simulations and test size 10,000 in three scenarios for a count outcome. For comparison, the value function of the true optimal regime, and the strategies of always treat and never treat are 3.36, 1.82, and 2.08, respectively.}
\centering
\begin{tabular}[t]{llrrrrrrrrrr}
\toprule
\multicolumn{2}{c}{} & \multicolumn{5}{c}{Scenario 1} & \multicolumn{5}{c}{Scenario 2}  \\
\cmidrule(l{3pt}r{3pt}){3-7} \cmidrule(l{3pt}r{3pt}){8-12} 
  &  & UA& ZZ1 & ZZ2& PDR1 & PDR2 & UA& ZZ1 & ZZ2 & PDR1 & PDR2  \\
\midrule
$n$=500 &&&&&&&&&\\
&ER & 0.13 &0.07&0.06 & 0.07 & 0.08 & 0.09&0.05 &0.06 &0.03 & 0.03 \\
&Value & 3.28& 3.34&3.35 & 3.34 & 3.33 &3.33 &3.35 &3.35 &3.36 & 3.36 \\
&FN & 0.00 & 0.00 &0.00& 0.00 & 0.00 & 0.00&0.00 & 0.00&0.00 & 0.00 \\
&FP & 1.00 &0.03&0.17 & 0.16 & 0.19 & 1.00& 0.00 &0.22 &0.04 & 0.01 \\

\midrule
$n$=1000 &&&&&&&&&\\
 &ER & 0.09& 0.06 & 0.05&0.04 & 0.04 & 0.06&0.04 & 0.05 &0.03 & 0.03 \\
&Value & 3.33& 3.35 & 3.35 &3.36 & 3.35 &3.35 &3.36 &3.36 & 3.36 & 3.36 \\
&FN & 0.00 &0.00& 0.00 &0.00 & 0.00&0.00 & 0.00 & 0.00&0.00 & 0.00 \\
&FP & 1.00& 0.00 & 0.16&0.07 & 0.08 & 1.00 &0.00 & 0.24&0.01 & 0.00  \\
\bottomrule
\end{tabular}
\label{tab:count_value}
\end{table}

Tables \ref{tab:count_value} and \ref{tab:binary_value} present the error rate (proportion of times the estimated optimal ITR fails to coincide with the true optimal ITR); value; false-negative rate (i.e., setting a tailoring variable's coefficient to 0 when it should be nonzero); false-positive rate (i.e., selecting a tailoring variable when the coefficient should in fact be zero) 
of the blip parameter estimates of the three methods for binary and count outcomes, respectively. In summary, all four methods have similar and good performance; this is expected as they are all doubly robust methods. For count outcomes, in scenario 1 with sample size 500, ZZ2 has the smallest error rate and the largest value; as the sample size increases to 1000, our proposed PDR1 outperforms other methods with respect to error rate and the value. As for scenario 2, our proposed PDR1 and PDR2 outperform all other competing methods regardless of the sample size. For example, when $n=1000$, PDR2 has the smallest error rate as well as the largest value; moreover, the FP and FN are both 0. The results of methods evaluated here for binary outcomes generally exhibit similarities to those for count outcomes.

\begin{table}[H]
\caption{Error rate (ER), value, false-negative (FN) and false-positive (FP) rate of variable selection results, with $n=500$ and 1000, for 400 simulations and a test size 10,000 in three scenarios for a binary outcome. For comparison, the value function of the true optimal regime, and the strategies of always treat and never treat are 0.64, 0.48, and 0.48, respectively.}
\centering
\begin{tabular}[t]{llrrrrrrrrrr}
\toprule
\multicolumn{2}{c}{} & \multicolumn{5}{c}{Scenario 1} & \multicolumn{5}{c}{Scenario 2}  \\
\cmidrule(l{3pt}r{3pt}){3-7} \cmidrule(l{3pt}r{3pt}){8-12} 
  &  & UA& ZZ1 & ZZ2& PDR1 & PDR2 & UA& ZZ1 & ZZ2 & PDR1 & PDR2  \\
\midrule
$n$=500 &&&&&&&\\
&ER & 0.18& 0.08 & 0.08 &0.07 & 0.07 &0.18 &0.07 &0.07 &0.07 & 0.08  \\
&Value & 0.61& 0.64 & 0.63 &0.64 & 0.64 & 0.61& 0.63 & 0.64 &0.64 & 0.64  \\
&FN & 0.00 & 0.00 & 0.00 &0.00 & 0.00 & 0.00& 0.00 & 0.00&0.00 & 0.00 \\
&FP & 1.00 &0.00 & 0.21 &0.05 & 0.05 & 1.00& 0.00 &0.21 &0.04 & 0.08 \\
\midrule
$n$=1000 &&&&&&&\\
&ER & 0.13& 0.07 & 0.07 &0.05 & 0.05 & 0.13 &0.06 & 0.05 &0.04 & 0.05 \\
&Value & 0.63& 0.64 & 0.64 &0.64 & 0.64 & 0.63&0.63 &0.64 &0.64 & 0.64 \\
&FN & 0.00 &0.00 & 0.00 &0.00 & 0.00 & 0.00&0.00 & 0.00 &0.00 & 0.00 \\
&FP & 1.00 &0.00 & 0.24 &0.03 & 0.02 & 1.00&0.00 & 0.25 &0.03 & 0.05 \\
\bottomrule
\end{tabular}
\label{tab:binary_value}
\end{table}

We make some final remarks on the simulation results here. First, no obvious difference in the error rate, value, and variable selection performance were observed between PDR1 and PDR2 in the simulations. Second, the penalization-based methods (PDRs and ZZ2) have a larger FP rate than the sequentially selection-based method ZZ1 in general. Specifically, ZZ1 has the best variable selection performance: for example, it achieves 0 FP rate for binary outcomes in both scenarios. Our proposed PDR approach has a slightly higher FP rate than ZZ1, however, it still can yield a larger value and a smaller error rate than ZZ1 in many settings. Moreover, our PDR approach has a much smaller FP rate than ZZ2, although they both are based on the $\ell_1$ penalty, PDR takes advantage of using the data-dependant adaptive weights and hence achieve a better variable selection performance than ZZ2.

In this section, we have focused exclusively on settings where assumptions are met. For a demonstration of the impact of violations of the assumption of correct specification of the blip model function, please see the Appendix C (Table C1). As anticipated, performance deteriorates significantly when this key assumption is not met.

\section{Application to an Adaptive Web-based Stress Management Study}

We illustrate the newly proposed approach on a dataset from a two-stage pilot of a sequential multiple assignment randomized trial \citep{lambert2021adaptive}. The trial aimed to assess a web-based, stress management intervention adapted across time using a stepped-care approach for people with cardiovascular disease. We focus our analysis on the first stage only, in which 50 participants were randomized into two treatment groups, each with probability $0.5$, stratified by recruitment source and stress level. The two treatment groups were: website only ($A=0$) and website plus weekly telephone coaching ($A=1$). 

The primary outcome in this analysis is the stress subscale from the Depression Anxiety Stress Scales (DASS) \citep{DASS}, which is a count outcome measured at 6 weeks after stage 1 randomized allocation. A lower DASS-stress subscale score suggests the presence of fewer symptoms of stress, so the optimal treatment decision minimizes the DASS-stress subscale score. The aims of our analysis were to determine the tailoring variables related to the decision rule and to obtain the estimated individualized treatment rule for individuals with cardiovascular disease. We restricted our analysis to eight variables: mental component score, age, DASS-stress subscale score at baseline, sex, marital status, stomach condition, physical component score, and vision. These were previously found to be useful for tailoring treatment using \citet{bian}.

A logistic regression model was posited to estimate the propensity score adjusted for the recruitment source and stress level. We applied PDR to this study with A-learning as the initial estimator (referred to as PDR1 in Section \ref{sim3}); both the baseline model and the blip model are posited to be linear. We found that five variables were relevant for tailoring treatment: DASS at baseline, sex, marital status, stomach condition, and vision. The estimated treatment rule is \begin{align*}
    \widehat a^{opt}=\mathbb{I}\{&-0.78+0.09\mathbb{I}(\mbox{male})+0.45\mathbb{I}(\mbox{unmarried})+0.01\mbox{DASS}+\\&0.45\mathbb{I}(\mbox{stomach=yes})-0.08\mathbb{I}(\mbox{vision=yes})<0\}.
\end{align*} For example, a married woman who does not have either a vision problem or a stomach ailment and who has a DASS greater than 13 would be recommended for website plus weekly telephone coaching ($A=1$). We compared our estimated treatment rule with results using the approach in \citet{bian}, treating the DASS as a continuous measure. We found that $74\%$ of subjects' recommended treatments were the same under the two strategies. Moreover, all five nonzero, estimated blip parameters had the same sign as the estimated blip parameters using \citet{bian}. 

We also considered, for illustrative purposes, an analysis that dichotomizes the outcome $Y$ at its median, using our proposed binary outcome approach. However, due to the small sample size, neither A-learning nor standard logistic regression yielded a solution, due to lack of convergence. 

Finally, we applied our newly proposed approach to data from the Sequenced Treatment Alternatives to Relieve Depression (STAR*D) study \citep{stard}. The STAR*D data are considered a benchmark dataset for ITR analyses and were analyzed in \citet{sb,al,modelg,bian}, among others. While these data are less novel, we considered the comparison relevant and provide results in the Appendix D. In summary, the findings in the current analysis, using the methods we propose for both count and binary outcomes, align well with the results found in \citet{sb,modelg,bian}.

\section{Discussion} \label{sec: discussion}

We proposed new, doubly robust estimating functions to estimate an ITR when the outcome is discrete and the log or logit link functions are used to model the outcome. The newly proposed approach can be solved using a weighted GLM iteratively, given a suitable choice of observational weights. The benefit of our proposed estimating function is that it is easily generalized to a penalized framework, which permits estimating a parsimonious ITR and selecting important tailoring variables simultaneously. Based on this finding, we also present a doubly robust criterion to select the tuning parameter. Numerical studies indicated that the newly proposed penalized doubly robust method compares favorably with other competing approaches in the context of ITRs. To our knowledge, doubly robust variable selection approach for ITRs with binary or count outcomes has not previously been studied.

We applied our proposed variable selection method to a sequential multiple assignment randomized trial \citep{lambert2021adaptive} to evaluate the effectiveness of a web-based stress management intervention for individuals with cardiovascular disease. We found that five variables were relevant for tailoring treatment: DASS at baseline, sex,
marital status, stomach condition, and vision. Furthermore, we derived a linear decision rule that may assist physicians in effectively recommending the web-based stress management intervention for patients with cardiovascular disease. This analysis yielded important insights into the influence (or lack thereof) of potential tailoring variables on patient primary outcomes, thus aiding to develop more effective and personalized approaches to care. 

One limitation of our proposed method is that we require that the parametric form of the blip function is known (i.e., that the blip is correctly specified). This requirement is slightly stronger than the assumption that the parametric form of the treatment regimes is correctly specified (see, e.g., \citet{zhang2022subgroup}), since assuming that the blip function is correct implies that the treatment regime is correct, but not the converse. An interesting avenue for future research would be to consider imposing smoothness assumptions on the blip function and estimating it using off-the-shelf non-parametric variable selection tools, for instance using splines with a penalty to control overfitting.

In this paper, for simplicity, we focus on a binary treatment setting. The extension to general discrete allocations, in which $a =\{0,1,..., l\}$, is straightforward: a multinomial model analogous to the generalized propensity score could be fit in place of $\pi$. Taking the outcomes to be counts, for example, the estimating function now is
 \begin{align*}
     &\sum_{a\neq 0} \sum_{i=1}^n\begin{pmatrix}\mathbb{I}(A_i=a)x^{\psi}_i\\ \xb_i \end{pmatrix}  |\mathbb{I}(A_i=a)- \mathbb P(A_i=a)| \exp\{-\gamma(\xp_i,a;\bpsi)\} \times \\&\left(y_i-\exp(f(\xb_i; \bbeta)+\gamma(\xp_i,a;\bpsi)\right)=0.
 \end{align*} As such, the estimation procedure and the theoretical results can be adapted without extra difficulty. Similarly, continuous exposure densities can be modelled directly, or approximated using quantile binning and modelled via a multinomial regression. Both of these approaches rely on a generalized propensity score \citep{imbens2000role} and were implemented in a continuous outcome setting for individualized treatment by \citet{SchulzMoodie2021}.

To obtain a doubly robust estimator, a well-behaved initial estimator is needed, which can be estimated using an unpenalized doubly robust approach. When the number of predictors is larger than the sample size, we recommend using the ridge estimator to acquire the initial estimate. In future work, we could also build on an idea in \citet{huang2008adaptive}, which used the marginal regression approach to obtain the initial estimator for the adaptive lasso (i.e., the outcome is regressed separately on each variable). However, this technique is more challenging in our setting, as it violates the assumption that the blip model is correctly specified. This is partial identification problem has been studied in \citet{van}, and this work may shed light on how to use marginal regression to obtain a valid initial estimator. It also may be of interest, in future work, to investigate the algorithm to directly solve the REE instead of using the approximation. As this alternative does not require an initial estimator, and it might perform better in a large $p$, small $n$ scenario. 

The extension of the single-stage estimation approach to a multistage setting also requires further investigation. In a multistage setting, the estimation procedure is conducted recursively using backward induction, and the ``outcome'' at each stage is set to be a predicted or estimated optimal response. For discrete outcomes, the optimal outcome is usually modeled by multiplicative effects, e.g., the optimal outcome at the $(k-1)$th stage for a count outcome is computed by $\widehat y_{k-1}^{opt}=y\times \prod_k^{K} \exp\{\gamma_k(\xp_k,\widehat a^{opt}_k;\bpsi_k)-\gamma_k(\xp_k,a_k;\bpsi_k)\}$, where $K$ is the total number of stages. A challenge under the multistage scenario is that the estimated optimal outcome at any stage for subjects with zero-valued outcome will always remain zero, unless adjustments are made \citep{modelg}, which may lead to a loss of efficiency.


\bibliographystyle{apalike}
\bibliography{bib}
\end{document}